\pdfoutput=1
\documentclass[fleqn,usenatbib, usedcolumn, letters]{mnras}

\usepackage[british]{babel}  
 
\usepackage{newtxtext,newtxmath}

\usepackage[T1]{fontenc}    
\usepackage{ae,aecompl}

\usepackage{graphicx}
\usepackage{amsmath,amssymb}

\title[Cresting the wave]{Cresting the wave: Proper motions of the Eastern Banded Structure} 
\author[Deason et al.]{
Alis J. Deason$^{1}$\thanks{E-mail: alis.j.deason@durham.ac.uk}, Vasily Belokurov$^{2}$, Sergey E. Koposov$^{2,3}$, \\
$^{1}$Institute for Computational Cosmology, Department of Physics, University of Durham, South Road, Durham DH1 3LE, UK\\
$^{2}$Institute of Astronomy, University of Cambridge, Madingley Road, Cambridge CB3 0HA, UK\\
$^{3}$McWilliams Center for Cosmology, Department of Physics, Carnegie Mellon University, 5000 Forbes Avenue, Pittsburgh, PA 15213, USA\\}
\date{Accepted XXX. Received YYY; in original form ZZZ}

\pubyear{2017}

\begin{document}

\label{firstpage}
\pagerange{\pageref{firstpage}--\pageref{lastpage}}

\maketitle

\begin{abstract}
We study the kinematic properties of the Eastern Banded Structure (EBS) and Hydra I overdensity using exquisite proper motions derived from the Sloan Digital Sky Survey (SDSS) and \textit{Gaia} source catalog. Main sequence turn-off stars in the vicinity of the EBS are identified from SDSS photometry; we use the proper motions and, where applicable, spectroscopic measurements of these stars to probe the kinematics of this apparent stream. We find that the EBS and Hydra I share common kinematic and chemical properties with the nearby Monoceros Ring. In particular, the proper motions of the EBS, like Monoceros, are indicative of prograde rotation ($V_\phi \sim 180-220$ km s$^{-1}$), which is similar to the Galactic thick disc. The kinematic structure of stars in the vicinity of the EBS suggest that it is not a distinct stellar stream, but rather marks the ``edge'' of the Monoceros Ring. The EBS and Hydra I are the latest substructures to be linked with Monoceros, leaving the Galactic anti-centre a mess of interlinked overdensities which likely share a unified, Galactic disc origin.

\end{abstract}


\section{Introduction}
Galaxies like the Milky Way consume and destroy several lower-mass dwarf galaxies over their lifetime. Evidence of these fatal accretion events can be seen in the form of streams, clouds, shells, and incoherent blobs that litter the stellar halo (e.g. \citealt{belokurov06}).  The interaction between accreted dwarf galaxies and the Galactic disc can not only rip apart the dwarf, but can perturb, warp, and even ``kick out'' stars from the stellar disc (e.g. \citealt{velazquez99, benson04, purcell10}). 

A stunning example of a substructure at the disc-halo interface, is the so-called Monoceros Ring. This local ($D\sim10$ kpc) overdensity located close to the Galactic anti-centre was first uncovered by \cite{newberg02} using imaging data from the Sloan Digital Sky Survey (SDSS; \citealt{york00}), and the ring-like structure was later confirmed  by \cite{ibata03}. There are two main formation scenarios that have been put forward to explain this vast anti-centre structure: (1) Tidal debris from a disrupted dwarf galaxy, and (2) a perturbation of the Galactic disc. Subsequent scrutiny of the anti-centre region has uncovered yet more substructure which could be related to Monoceros: the Canis Major overdensity, which has been touted as a potential progenitor of Monoceros (\citealt{martin04}),  the Eastern Banded Structure (EBS), the anti-centre stream \citep{grillmair06}, and a Southern counterpart to Monoceros at larger distances ($D > 15$ kpc) called TriAnd (\citealt{majewski04, rochapinto04}). More recently it has been suggested that the disc in this region has complex substructure, which is caused by an accreted dwarf galaxy. Thus, these apparently spatially separated substructures could have a unified origin (see e.g. \citealt{slater14, xu15}).
\begin{figure*}
    \centering
    \includegraphics[width=14cm, height=9.33cm]{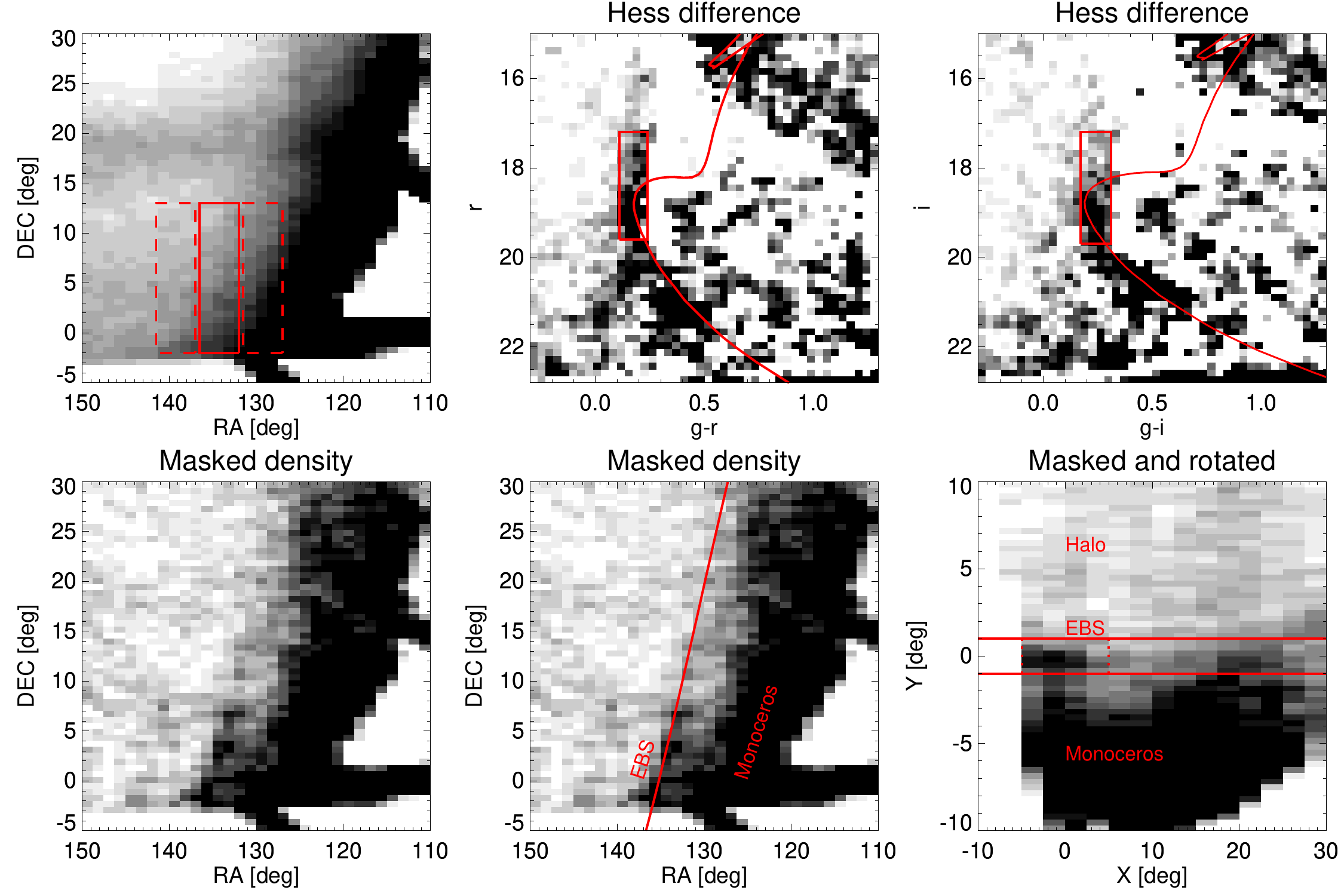}
    \caption[]{\textit{Top left panel:} Density map of SDSS
      stars in Equatorial coordinates. Only stars with main sequence colors
    ($0.2 < g-r < 0.6$) in the magnitude range $17 < r < 22$ are
    shown. The red box indicates the EBS,
    and the dashed boxes show the regions used to calculate the
    background. \textit{Top middle and right panels:}
    Hess diagrams in $g-r$, $r$ (middle) and $g-i$, $i$ (right)
    color-magnitude space. Here, we show the difference between stars
    in the EBS region and the background. The red box indicates the
    turn-off feature associated with the EBS, and the red line shows a 
  5.6 Gyr, [Fe/H]=-0.9 isochrone at $D=12.7$ kpc. \textit{Bottom left and middle panels:} The
  density of turn-off stars selected in the color and magnitude
  ranges: $0.11 < g-r < 0.24$, $17.2 < r < 19.6$, $0.17 < g-i < 0.31$,
  $17.2 < i < 19.7$. The red line in the bottom middle panel shows a great circle through the EBS. \textit{Bottom right panel:} Density of
  turn-off stars in a coordinate system with equator aligned with the
  EBS stream. Hydra I lies at $X,Y=[0,0]$ deg. The solid red
  line indicates the approximate boundary of the EBS stream, $|Y| < 1.0$ deg, and
  the dotted line shows the range along the stream
  where there is obvious EBS signal ($|X| < 5$ deg).}
    \label{fig:ebs_sel}
\end{figure*}
In this letter, we focus on the EBS, a faint, short structure first discovered by \cite{grillmair06}. Despite its proximity to Monoceros, a re-analysis by \cite{grillmair11} found that the EBS was unlikely related to the Ring, and is instead a distinct stellar stream. Indeed,  \cite{grillmair11} suggested that a relatively high surface density, double-lobed feature found along the stream - called Hydra I - could be the dwarf galaxy or globular cluster progenitor of the stream. However, while the shallow SDSS imaging assumed an old, metal-poor stellar population, deeper imaging and spectroscopy of Hydra I by \cite{hargis16} revealed that this overdensity is of intermediate age (5-6 Gyr) and metallicity ([Fe/H] $= -0.9$). This relatively high metallicity is reminiscent of the Monoceros Ring ([Fe/H] $\sim-1.0$, e.g \citealt{ivezic08, meisner12}), which is only slightly more metal-poor than the thick disc ([Fe/H] $\sim -0.7$, e.g \citealt{gilmore85}).

It is clear that a kinematic exploration of the anti-centre region is desperately needed to dissect these complex structures. To this end, \cite{deboer17} recently exploited a proper motion catalog constructed from the SDSS and first data release of \textit{Gaia} to infer the tangential motions of the Monoceros Ring. In this letter, we use the same SDSS-\textit{Gaia} catalog to examine the kinematics of the EBS and Hydra I, and relate them to the Ring.

\section{SDSS-\textit{Gaia} Proper Motions}
\label{sec:pms}

In order to examine the proper motions of the EBS we use a newly calibrated SDSS-\textit{Gaia} catalog (Koposov et al. in prep).
The details of the creation of the recalibrated SDSS astrometric catalogue, the measurement of SDSS-\textit{Gaia} proper
motions, and the statistical and systematic uncertainties of the derived proper motions, are described in more detail in Koposov et al. (in prep), \cite{deason17} and \cite{deboer17}. Here, we give a very brief description.

A comparison between the positions of stars in the SDSS Data Release 10 \citep{ahn14} and the \textit{Gaia} source catalog \citep{gaia_dr1} with a baseline of $\sim5-10$ years can potentially yield proper motions across a large area of sky. However, there are several systematics present in this naive cross-match, particularly due to the astrometric solution of the SDSS survey. To this end, Koposov et al. (in prep) recalibrate the SDSS astrometry based on the excellent astrometry of the \textit{Gaia} source catalog. These recalibrated positions are then cross-matched with the \textit{Gaia} catalog resulting in proper motion measurements for the majority of sources down to $r \sim 20$ mag. The precision of the new SDSS-\textit{Gaia} proper motions are examined in \cite{deason17} and \cite{deboer17} using spectroscopically confirmed QSOs. The systematics of the catalog are very low ($\sim 0.1$ mas/yr) and the statistical uncertainties are typically 1.5 mas/yr. In this work, we adopt the proper motion uncertainties as a function of magnitude derived by \cite{deboer17} for the anti-centre region: $\sigma_{\mu_l}=36.04-3.867g+0.107 g^2$, $\sigma_{\mu_b}=26.50-2.894 g+0.082 g^2$.

\section{Sample Selection}
\label{sec:samples}

\begin{figure*}
    \centering
    \includegraphics[width=14cm, height=4.66cm]{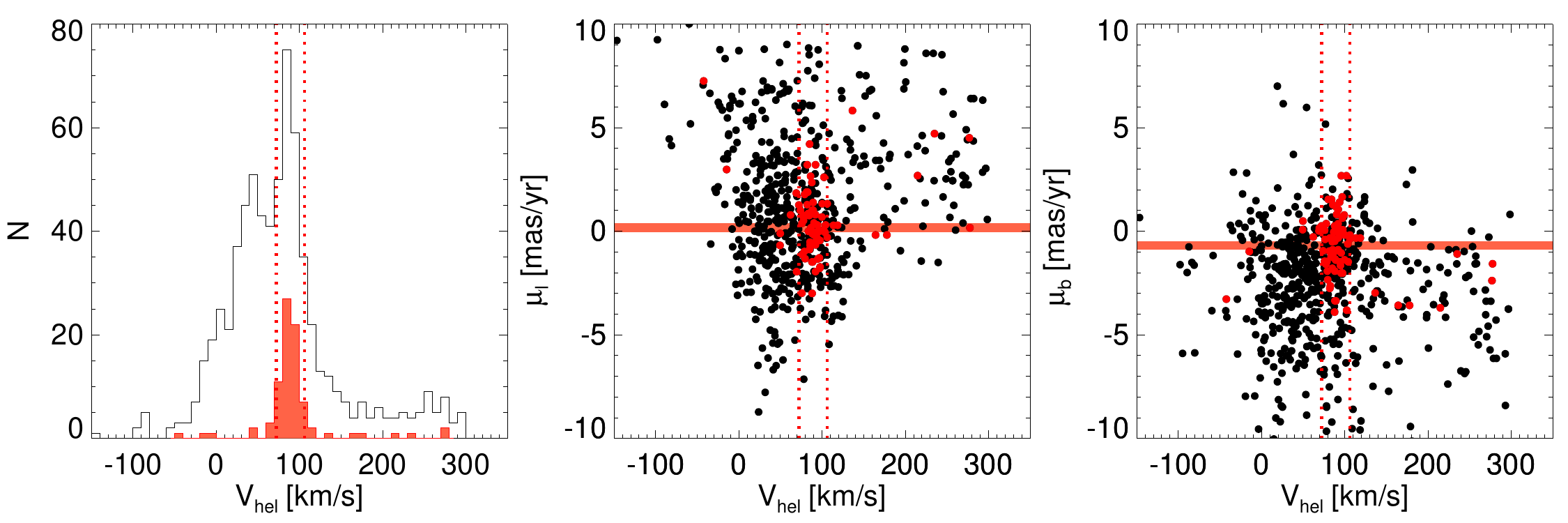}
    \caption[]{Sample of $N=675$ candidate Hydra I stars in the
      SDSS-\textit{Gaia} catalog with spectroscopic
      measurements from \cite{hargis16} (H16). \textit{Left panel:}
      Distribution of heliocentric velocities of the H16 sample (black histogram). The
      shaded red histogram shows the $N=84$ stars that lie within our
      color and magnitude selection boundaries. These stars have a narrowly peaked velocity distribution with median $V_{\rm hel} = 89$ km
      s$^{-1}$, in excellent agreement with H16. For our proper motion measurements we use the
    $N=65$ stars that have heliocentric velocities within 2$\sigma_V=2 \times 8.4$ km s$^{-1}$ of
  the average Hydra I velocity. \textit{Middle and right panels:} The Galactic
  proper motions of the H16 sample as a function of heliocentric
  velocity. The stars that pass our turn-off selection cut are shown
  in red. The median proper motions for the turn-off stars with
  $V_{\rm hel}$ within $2 \sigma_v$ of the net Hydra I velocity are
  shown with the red shaded regions. Here, the width of the lines
  indicate the $1 \sigma$ errors of the medians.}
    \label{fig:hyd_pm}
\end{figure*}

In this Section, we outline our selection of stars associated with the EBS. In Fig. \ref{fig:ebs_sel} we show how EBS stars are identified according to their position on the sky, color and magnitude. The top left panel shows a density map of SDSS stars in the vicinity of the EBS. Here, only stars with main sequence colors ($0.2 < g-r < 0.6$) in the magnitude range $17 < r < 22$ are shown. The high density of stars at RA $\sim 110-130$ deg is due to the Monoceros overdensity. The solid red box indicates the region of the EBS, and the dashed red boxes show the background regions we use to make differential Hess diagrams. The Hess difference between the EBS and the background is shown in the middle and right panels. We indicate our selection of main sequence turn-off stars with the red box. There is an obvious signal of turn-off stars in the EBS relative to the background. \cite{hargis16} (H16 hereafter) use deeper imaging with DECam to show that Hydra I has a well-defined main sequence turn-off of intermediate age ($\sim 5-6$ Gyr) and metallicity ([Fe/H] =-0.9). Indeed, the red line shows that a 5.6 Gyr, [Fe/H]=-0.9 isochrone at the distance of Hydra I ($D=12.7$ kpc) agrees well with the turn-off signal in the EBS. In the bottom left and middle panels we show the density of stars on the sky, which are selected to have the same color and magnitude as the EBS turn-off. These color and magnitude constraints are given by:
\begin{align}
0.11 < g -r < 0.24 \notag\\
17.2 < r < 19.6 \notag \\
0.17 < g-i < 0.31 \notag \\
17.2 < i < 19.7
\label{eq:to}
\end{align}
The EBS now becomes clear in these panels, and the red line in the bottom middle panel shows an approximate great circle through the EBS (with pole at RA, DEC= [-134.5, 14.8] deg). In the bottom right panel we transform to a coordinate system aligned with this great circle, with Hydra I at the origin ($X, Y = [0,0]$ deg). Here, the EBS signal is apparent at $|X| < 5$ deg along the stream and is approximately confined to $|Y| < 1.0$ deg. 

In the following sections, we use turn-off stars selected according to Eqn. \ref{eq:to} to identify EBS stars and use the rotated ($X, Y$) coordinate system to dissect this structure in proper motion space.

\section{Proper Motion of Hydra I}
\label{sec:hydra}

Before we turn our attention to the full extent of the EBS, we focus on the potential progenitor of this stream, Hydra I. Here, we use the spectroscopic sample of stars associated with Hydra I compiled by H16, and cross-match these stars with our SDSS-\textit{Gaia} proper motion catalog. The top left panel of Fig. \ref{fig:hyd_pm} shows the distribution of heliocentric velocities of the $N=675$ candidate Hydra I stars. The filled red histogram shows the $N=84$ turn-off stars that lie within the color and magnitude selection boundaries given in Eqn. \ref{eq:to}.

H16 find a (probability weighted) systemic velocity of $V_{\rm hel} = 89.4$ km s$^{-1}$ for Hydra I, with dispersion $\sigma_V = 8.4$ km s$^{-1}$. Our turn-off star criteria produces a well defined peak centered on this systemic velocity, and thus is clearly effective at isolating stars associated with Hydra I. The dotted lines in this panel indicate the 2$\sigma$ boundaries of the Hydra I heliocentric velocity distribution.

The middle and right panels of Fig. \ref{fig:hyd_pm} show the Galactic proper motion components of the candidate Hydra I stars. Stars that fall into our turn-off selection are highlighted in red, and the dotted red lines indicate the $2\sigma_v$ boundary of the systemic velocity of Hydra I. Using the $N=65$ turn-off selected stars that have heliocentric velocities within $2\sigma_v$ of Hydra I, we find median proper motions of: $\mu_l = 0.18 \pm 0.21$ mas/yr and $\mu_b =-0.67 \pm 0.21$ mas/yr. The median values, and uncertainties in the medians, are calculated using a bootstrap method that takes into account the proper motion uncertainties of the SDSS-\textit{Gaia} sample. 

We can convert our derived proper motions and the line-of-sight velocity of Hydra I into a 3D velocity. Heliocentric quantities are converted to Galactocentric ones using a solar motion of $v_{\phi, \odot}=238$ km s$^{-1}$ and $(U,V,W)=(11.1,12.24,7.25)$ km s$^{-1}$ from \cite{reid04} and \cite{schonrich10}. Adopting a distance to Hydra I of 12.7 kpc, we find: $(v_r, v_\theta, v_\phi)=(-22 \pm 5, -24 \pm 12, 179 \pm 12)$ km s$^{-1}$. Thus, the kinematics of Hydra I indicate a prograde orbit, with similar amplitude to the Galactic thick disc.

\section{Kinematics of the Eastern Banded Structure}
\label{sec:ebs}
\subsection{Proper motion properties}
\begin{figure}
  \begin{center}
    \begin{minipage}{\linewidth}
      \centering
     \includegraphics[width=8.5cm, height=4.25cm]{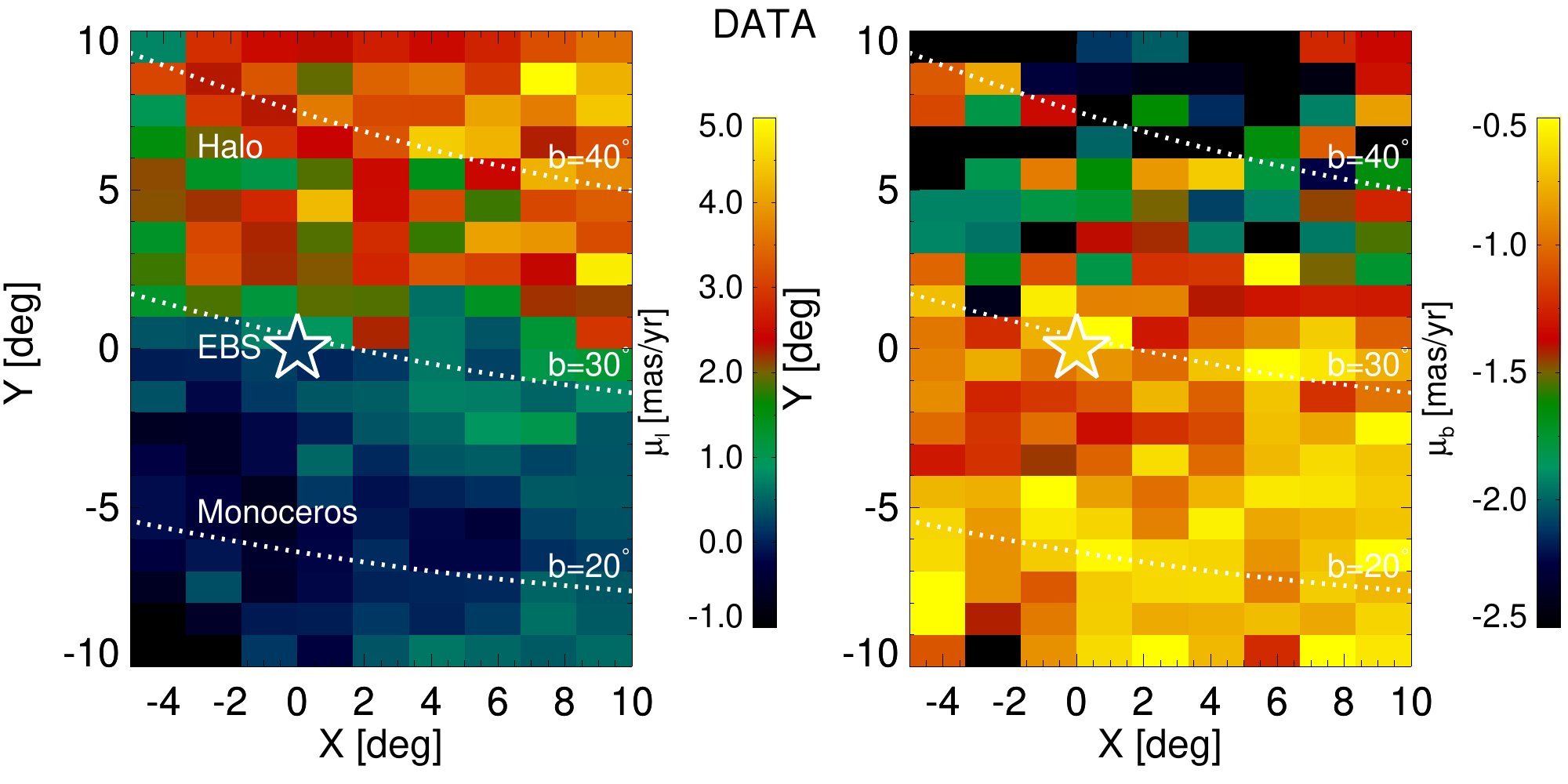}
   \end{minipage}
     \begin{minipage}{\linewidth}
       \includegraphics[width=8.5cm, height=4.25cm]{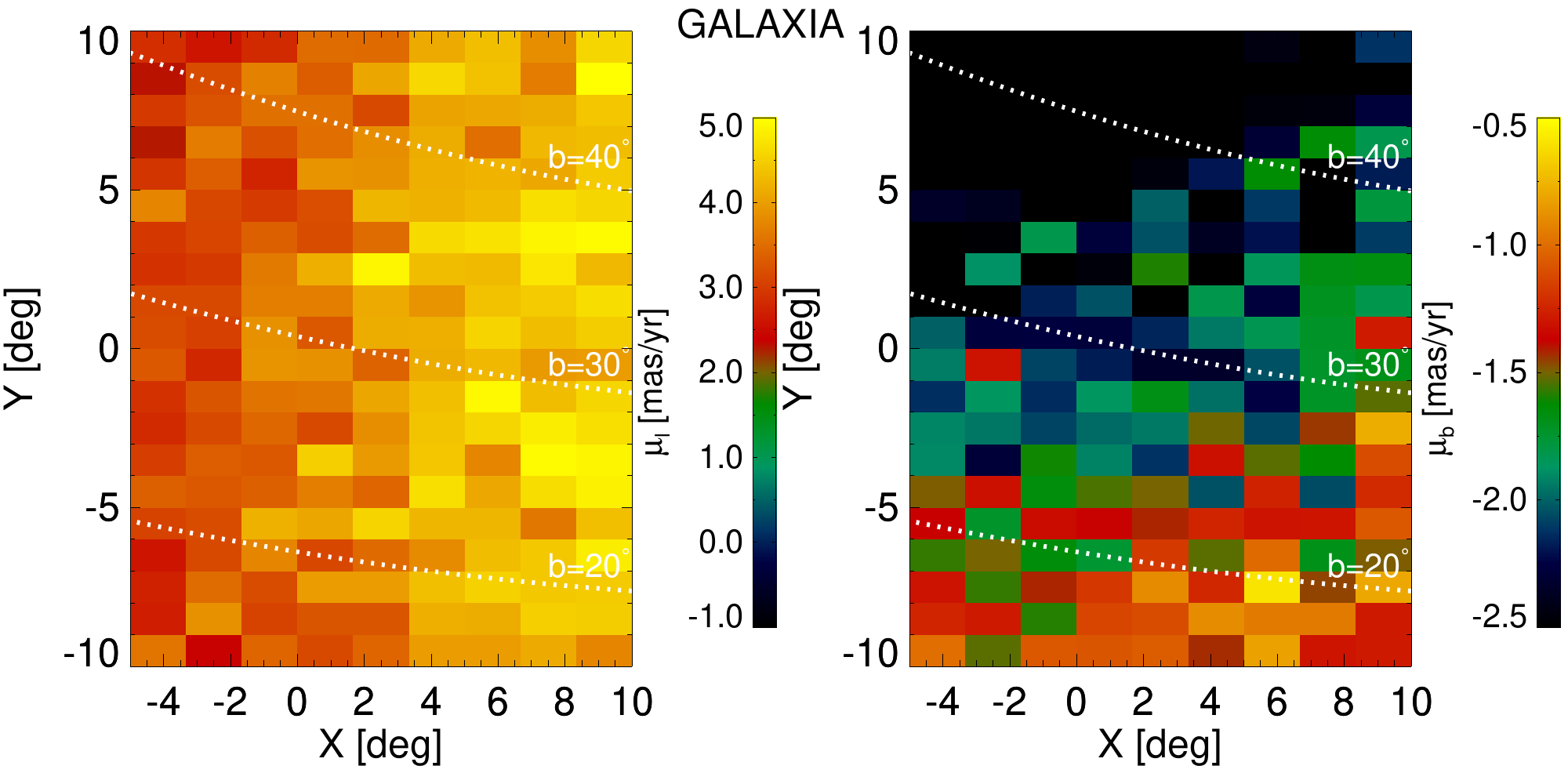}
     \end{minipage}
     \caption[]{\textit{Top panels:} The median proper motions of turn-off stars in
       the vicinity of the EBS. The coordinate
       system has equator aligned with the EBS, and we show lines of constant Galactic latitude. The density of EBS
       stars is strongest in the region $|Y| < 1.0$ deg and $|X| < 5$
       deg. The left and right panels show the Galactic longitude and
       latitude proper motion components, respectively. At negative
       $Y$ the Monoceros Ring dominates, and at positive $Y$ the halo
       population becomes more apparent. The EBS is at the ``edge'' of
       Monoceros and its proper motions indicate a very similar
       kinematic signature to the Monoceros Ring. The color of the star symbols
       indicate the proper motion of Hydra I calculated in Section
       \ref{sec:hydra}. The \textit{bottom
       panels} show the proper motion distributions for the Galaxia
       model, where stars have been selected using the same color and
       magnitude cuts. Note the gradient in $\mu_b$ along Y in Galaxia is largely due to solar reflex motion. The rotation component is mainly in the $\mu_l$ direction hence this coordinate is more discriminating between rotation and non-rotation than $\mu_b$ or $V_{\rm hel}$.}
    \label{fig:ebs_pm_xy}
   \end{center}
\end{figure}

\begin{figure}
  \begin{center}
    \begin{minipage}{0.48\linewidth}
      \centering
     \includegraphics[width=4.49cm, height=7.7cm]{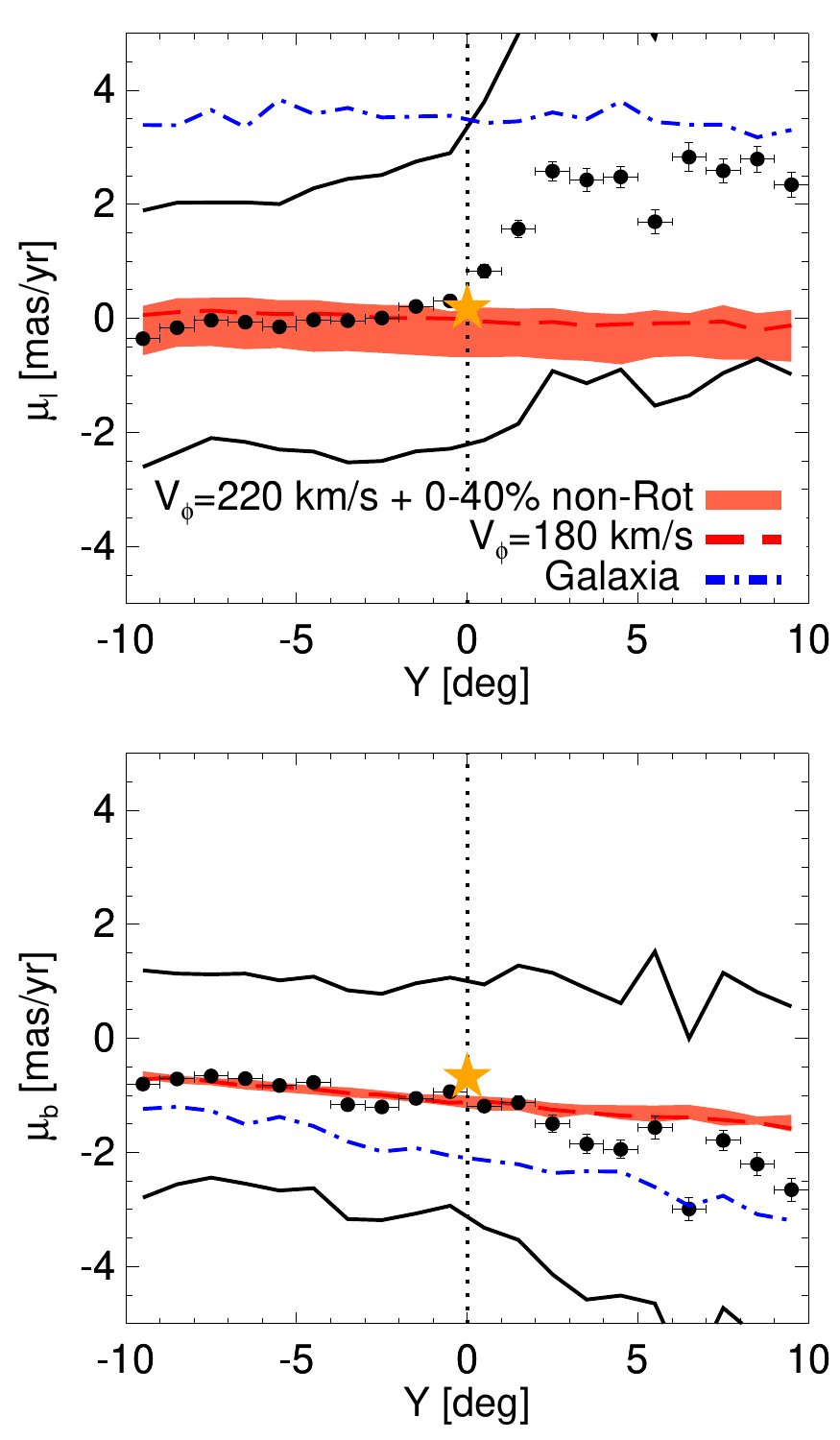}
    \end{minipage}
    \hspace{1.5pt}
     \begin{minipage}{0.48\linewidth}
       \includegraphics[width=4.11cm, height=7.7cm]{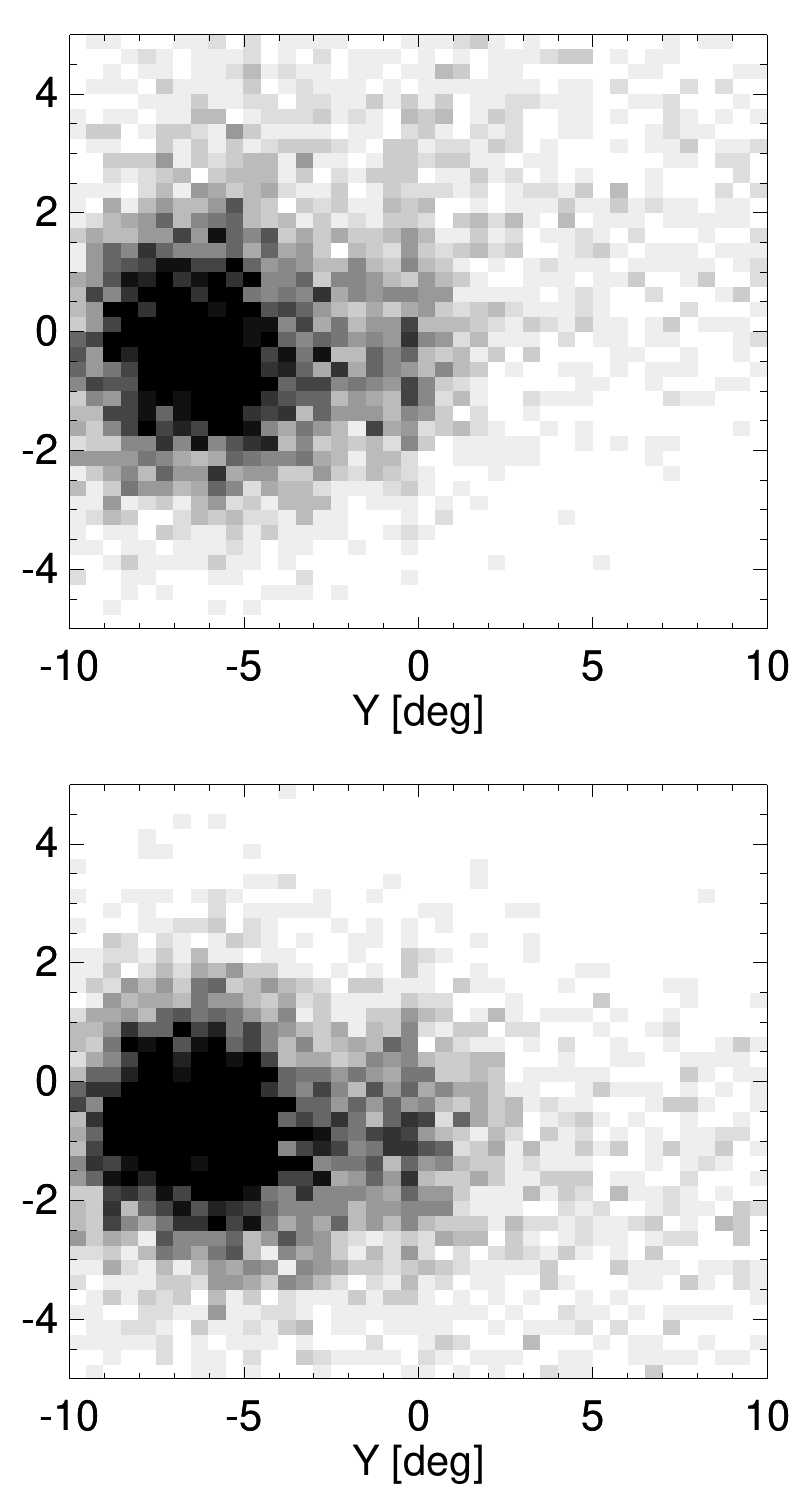}
     \end{minipage}
    \caption[]{\textit{Left panels:} Galactic proper motions along the $Y$ coordinate
      perpendicular to the EBS stream direction. We only use stars with $1 < |X|/\mathrm{deg} < 5$. The black points show the
    median values and the solid lines indicate the $1 \sigma$
    dispersion. The horizontal and vertical error bars on the black points indicate the binsize and error in the median, respectively. The yellow star indicates the values for Hydra I. The
    proper motions of stars in the EBS (at $Y \sim 0$ deg) closely follow
    the kinematic signature of the Monoceros Ring, which dominates at
    $Y < 0$ deg. The vertical dotted black line indicates the Monoceros/EBS to
    halo transition, and the dot-dashed blue line shows the predictions of the
  Galaxia model. The dashed red line indicates a simple disc model with $V_\phi = 180$ km s$^{-1}$ at $D=12.7$ kpc. The solid red region indicates a disc model with $V_\phi = 220$ km s$^{-1}$, which includes a fraction of non-rotating (i.e. halo) stars that varies from 0-40\%. \textit{Right panels:} Density map of proper motions as a function of Y.}
    \label{fig:ebs_pm_y}
    \end{center}
\end{figure}

We now examine the proper motions of stars in the vicinity of the EBS. Here, we only use turn-off stars (see Fig. \ref{fig:ebs_sel} and Eqn. \ref{eq:to}) and use a rotated coordinate system aligned with a great circle through the EBS. In the top panels of Fig. \ref{fig:ebs_pm_xy} we show the median Galactic longitude (left panels) and latitude (right panels) proper motions of stars close to the EBS in bins of $X$ and $Y$. Here, Hydra I is located at the origin ($|X| < 1$ deg, $|Y| < 1$ deg), and the star symbol indicates its estimated proper motion found in the previous section. For comparison, we show the proper motions of stars in the Galaxia model \citep{sharma11} in the bottom panels. These stars are selected according to the same criteria as the EBS stars, and the model is dominated by halo stars in this region of the sky.

The proper motion structure in the vicinity of the EBS shows a sharp transition at $Y \sim 0$ deg that is not apparent in the Galaxia model. The Y coordinate runs perpendicular to the stream, and the Monoceros Ring dominates at $Y < 0$ deg. Thus, the EBS seems to mark the ``edge'' of the Monoceros feature, and indeed the EBS proper motions (at $|Y| < 1.0$ deg and $1 < |X|/\mathrm{deg} < 5$) are very similar to Monoceros. In Fig. \ref{fig:ebs_sel} the density of the anti-centre features vary considerably with Galactic latitude (i.e. the EBS stands out as an obvious overdensity). Thus, if the EBS was a distinct structure, we would also expect to see a variation, or discontinuity, in the proper motions. Instead, we find that the proper motions in the vicinity of the EBS follow on continuously from the Monoceros Ring, as would be expected if the Monoceros star counts continue to dominate out to the EBS.

The transition between Monoceros and the halo can be seen more clearly in Fig. \ref{fig:ebs_pm_y}, where we collapse the $X$ dimension (along the stream) and only use stars with $1 < |X| < 5$ deg. Note that we restrict to $|X| > 1$ deg to exclude Hydra I stars. Here, we show how the Galactic proper motion components vary perpendicular to the stream. The filled points show the median values (with associated errors) and the solid lines indicate the 1$\sigma$ dispersion. The orange star symbol indicates Hydra I, and the dot-dashed blue line shows the prediction from the Galaxia model. The solid red region shows a simple disc model with rotation $V_\phi = 220$ km s$^{-1}$ at $D=12.7$ kpc, where we assume the distance to EBS is the same as Hydra I. We have also included a non-rotating component, which contributes between 0-40\% to the model. Here, we assume a range of distances between 5-100 kpc that follow a power-law density distribution with slope $-3.5$.  Note that \cite{deboer17} found that contamination from the Milky Way varies between 20-40\% in the region of the sky that is dominated by Monoceros.
 
 The model with prograde rotation $V_\phi= 220$ km s$^{-1}$ at $D=12.7$ kpc follows the Monoceros stars, in good agreement with the prograde models derived by \cite{deboer17} for the Ring in this region of the sky (see Figures 10 and 11 in \citealt{deboer17}). However, the stars nearby to the EBS also lie close to this prograde model. Indeed, it is only at  $Y \gtrsim 2$, beyond the EBS, that the proper motions transition to a more halo-like population.
 
\subsection{Metallicity and line-of-sight velocity}
\begin{figure}
    \centering
    \includegraphics[width=7cm, height=10.5cm]{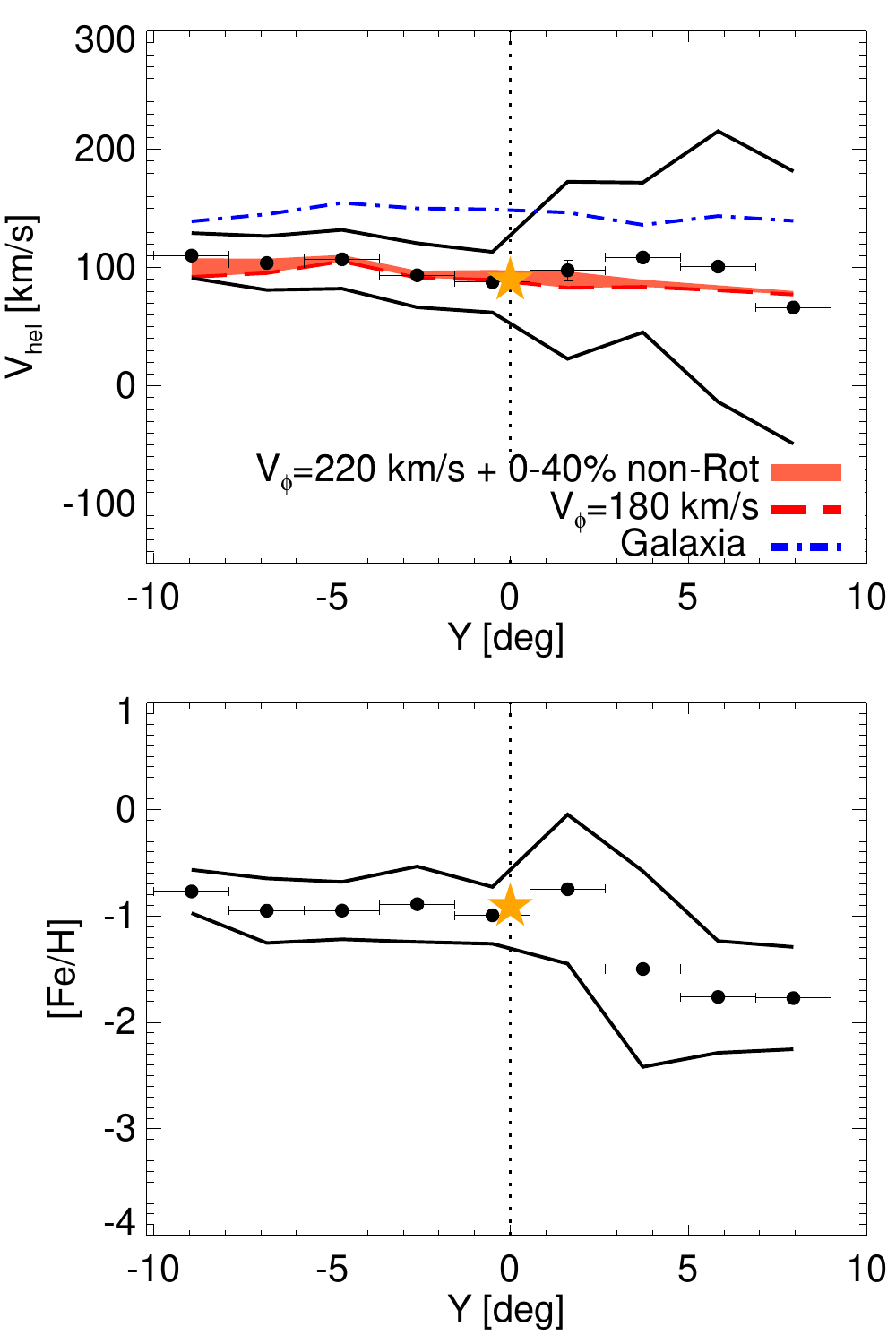}
    \caption[]{The heliocentric velocity (top panel) and
      metallicity (bottom panel) of stars in the vicinity of the
      EBS. Here, stars in SDSS with spectroscopy have been selected
      according to the color and magnitude boundaries described in
      Fig. \ref{fig:ebs_sel}. The yellow stars indicate the values for Hydra I derived by \cite{hargis16}. The EBS at Y $\sim 0$ deg has similar
      heliocentric velocity and metallicity as the Monoceros Ring. When
    $Y > 0$ deg, the distribution transitions to the metal-poor
    halo. The dotted black line indicates the Monoceros/EBS to halo transition.}
    \label{fig:ebs_spec}
\end{figure}

For our final kinematic examination of the EBS, we isolate turn-off stars in the vicinity of the EBS with spectroscopic measurements from SDSS. Here, we adopt the spectroscopic parameters derived from the Sloan Extension for Galactic Exploration and Understanding (SEGUE, \citealt{yanny09}) Stellar Parameter Pipeline (SSPP, \citealt{lee08}). In the region $|X| < 10$, $|Y| < 10$ there are $N=817$ stars with line-of-sight velocity and metallicity measurements.  In Figure \ref{fig:ebs_spec} we show the median heliocentric velocity (top panel) and metallicity (bottom panel) in bins perpendicular to the stream (note only stars with $|X| < 5$ deg are used). Similarly to Fig. \ref{fig:ebs_pm_y}, the spectroscopic properties of the EBS are similar to the Monoceros Ring, which dominates the $Y <0$ deg star counts. The heliocentric velocity of the EBS is indicative of a prograde rotating model with $V_\phi \sim 180- 220$ km s$^{-1}$. In addition, the EBS stars are relatively metal-rich ([Fe/H] $\sim -0.9$) in good agreement with what H16 found for Hydra I. It is only beyond the EBS ($ Y \gtrsim 2$ deg) that the metallicity more closely resembles a metal-poor halo population.

\section{Summary and Discussion}
In this letter, we have examined the kinematic properties of the EBS and Hydra I using exquisite proper motions derived from the SDSS and \textit{Gaia} source catalog. We find that the EBS and Hydra I share common kinematic and chemical properties with the Monoceros Ring; the proper motions, line-of-sight velocity, distance and metallicity of the EBS \textit{all} closely resemble Monoceros and do not appear like a halo population. In particular, the proper motions of the EBS, like Monoceros, are indicative of prograde rotation, which only slightly lags behind the Galactic thin disc. At higher Galactic latitudes than the EBS, the stellar field transitions to a halo dominated population. Thus, we conclude that the EBS is not a distinct stellar stream, but rather marks the ``edge'' of the Monoceros Ring.

A detailed kinematic analysis of the Galactic anti-centre was recently performed by \cite{deboer17}. These authors used the same SDSS-\textit{Gaia} proper motion catalog described in this work, and find that the Monoceros Ring has a prograde rotation signal with $V_\phi \sim 220-240$ km s$^{-1}$. \cite{deboer17} also map the proper motion structure of the anti-centre stream (ACS), and although they find that it is distinct from Monoceros, the ACS has similar kinematic properties and likely shares a common origin with the Ring. There are two leading formation mechanisms for the Monoceros structure: (1) a disrupted dwarf galaxy and (2) a perturbation of the Galactic disc. \cite{deboer17} suggest that their results indicate a mix of these two scenarios, where both are likely needed to explain the complex structures in the anti-centre. 

Our results point to yet another Galactic sub-structure associated with the Monoceros Ring. Indeed, the kinematic structure of the EBS could be likened to the ``crest'' of the Monoceros wave. We also find that Hydra I, the presumed progenitor of the EBS, is both kinematically and chemically related to both the EBS and Monoceros. The fact that these substructures, in addition to the ACS, are not substantially different from Monoceros puts the disrupted dwarf scenario into serious doubt. Indeed, the main remaining evidence for the debris of an accreted satellite are the reported detections of thin substructures in the anti-centre region (like EBS). Thus, our finding that the EBS is in fact not a distinct stream, and is instead an overdensity related to Monoceros, indicates that the similarity of the various stellar components in the anti-centre favors a unified, disc origin.

The Galactic anti-centre is proving to be a region of the Galaxy littered with multiple, interlinked sub-structures. This unique probe of the disc/halo interface not only allows us to study the disruption of dwarf galaxies, but also the synergy between accretion events and the response of the Galactic disc. With the advent of more detailed kinematics from \textit{Gaia} and upcoming spectroscopic surveys (such as DESI and WEAVE; \citealt{desi,weave}), we can hope to piece together the complex formation scenario of Monoceros and its denizens.

\label{lastpage}

 \section*{acknowledgments}
We thank an anonymous referee for a thorough report, which improved the clarity of this manuscript.
A.D. is supported by a Royal Society University Research Fellowship. 
The research leading to these results has received funding from the
European Research Council under the European Union's Seventh Framework
Programme (FP/2007-2013) / ERC Grant Agreement n. 308024. V.B. and S.K. acknowledge financial support from the ERC. A.D. and S.K. also acknowledge the support from the STFC (grants ST/P000541/1 and ST/N004493/1).

\bibliographystyle{mnras}

\begin{thebibliography}{}
\makeatletter
\relax
\def\mn@urlcharsother{\let\do\@makeother \do\$\do\&\do\#\do\^\do\_\do\%\do\~}
\def\mn@doi{\begingroup\mn@urlcharsother \@ifnextchar [ {\mn@doi@}
  {\mn@doi@[]}}
\def\mn@doi@[#1]#2{\def\@tempa{#1}\ifx\@tempa\@empty \href
  {http://dx.doi.org/#2} {doi:#2}\else \href {http://dx.doi.org/#2} {#1}\fi
  \endgroup}
\def\mn@eprint#1#2{\mn@eprint@#1:#2::\@nil}
\def\mn@eprint@arXiv#1{\href {http://arxiv.org/abs/#1} {{\tt arXiv:#1}}}
\def\mn@eprint@dblp#1{\href {http://dblp.uni-trier.de/rec/bibtex/#1.xml}
  {dblp:#1}}
\def\mn@eprint@#1:#2:#3:#4\@nil{\def\@tempa {#1}\def\@tempb {#2}\def\@tempc
  {#3}\ifx \@tempc \@empty \let \@tempc \@tempb \let \@tempb \@tempa \fi \ifx
  \@tempb \@empty \def\@tempb {arXiv}\fi \@ifundefined
  {mn@eprint@\@tempb}{\@tempb:\@tempc}{\expandafter \expandafter \csname
  mn@eprint@\@tempb\endcsname \expandafter{\@tempc}}}

\bibitem[\protect\citeauthoryear{{Ahn} et~al.,}{{Ahn} et~al.}{2014}]{ahn14}
{Ahn} C.~P.,  et~al., 2014, \mn@doi [\apjs] {10.1088/0067-0049/211/2/17}, \href
  {http://adsabs.harvard.edu/abs/2014ApJS..211...17A} {211, 17}

\bibitem[\protect\citeauthoryear{{Belokurov} et~al.,}{{Belokurov}
  et~al.}{2006}]{belokurov06}
{Belokurov} V.,  et~al., 2006, \mn@doi [\apjl] {10.1086/504797}, \href
  {http://adsabs.harvard.edu/abs/2006ApJ...642L.137B} {642, L137}

\bibitem[\protect\citeauthoryear{{Benson}, {Lacey}, {Frenk}, {Baugh}  \&
  {Cole}}{{Benson} et~al.}{2004}]{benson04}
{Benson} A.~J.,  {Lacey} C.~G.,  {Frenk} C.~S.,  {Baugh} C.~M.,   {Cole} S.,
  2004, \mn@doi [\mnras] {10.1111/j.1365-2966.2004.07870.x}, \href
  {http://adsabs.harvard.edu/abs/2004MNRAS.351.1215B} {351, 1215}

\bibitem[\protect\citeauthoryear{{de Boer}, {Belokurov}  \& {Koposov}}{{de
  Boer} et~al.}{2017}]{deboer17}
{de Boer} T.~J.~L.,  {Belokurov} V.,   {Koposov} S.~E.,  2017, preprint, \href
  {http://adsabs.harvard.edu/abs/2017arXiv170609468D} {} (\mn@eprint {arXiv}
  {1706.09468})
  
\bibitem[\protect\citeauthoryear{{DESI Collaboration} et~al.,}{{DESI
  Collaboration} et~al.}{2016}]{desi}
{DESI Collaboration} et~al., 2016, preprint, \href
  {http://adsabs.harvard.edu/abs/2016arXiv161100036D} {} (\mn@eprint {arXiv}
  {1611.00036})

\bibitem[\protect\citeauthoryear{{Dalton} et~al.,}{{Dalton}
  et~al.}{2012}]{weave}
{Dalton} G.,  et~al., 2012, in Ground-based and Airborne Instrumentation for
  Astronomy IV. p. 84460P, \mn@doi{10.1117/12.925950}

\bibitem[\protect\citeauthoryear{{Deason}, {Belokurov}, {Koposov}, {G{\'o}mez},
  {Grand}, {Marinacci}  \& {Pakmor}}{{Deason} et~al.}{2017}]{deason17}
{Deason} A.~J.,  {Belokurov} V.,  {Koposov} S.~E.,  {G{\'o}mez} F.~A.,  {Grand}
  R.~J.,  {Marinacci} F.,   {Pakmor} R.,  2017, \mn@doi [\mnras]
  {10.1093/mnras/stx1301}, \href
  {http://adsabs.harvard.edu/abs/2017MNRAS.470.1259D} {470, 1259}

\bibitem[\protect\citeauthoryear{{Gaia Collaboration} et~al.,}{{Gaia
  Collaboration} et~al.}{2016}]{gaia_dr1}
{Gaia Collaboration} et~al., 2016, \mn@doi [\aap]
  {10.1051/0004-6361/201629512}, \href
  {http://adsabs.harvard.edu/abs/2016A%26A...595A...2G} {595, A2}

\bibitem[\protect\citeauthoryear{{Gilmore} \& {Wyse}}{{Gilmore} \&
  {Wyse}}{1985}]{gilmore85}
{Gilmore} G.,  {Wyse} R.~F.~G.,  1985, \mn@doi [\aj] {10.1086/113907}, \href
  {http://adsabs.harvard.edu/abs/1985AJ.....90.2015G} {90, 2015}

\bibitem[\protect\citeauthoryear{{Grillmair}}{{Grillmair}}{2006}]{grillmair06}
{Grillmair} C.~J.,  2006, \mn@doi [\apjl] {10.1086/509255}, \href
  {http://adsabs.harvard.edu/abs/2006ApJ...651L..29G} {651, L29}

\bibitem[\protect\citeauthoryear{{Grillmair}}{{Grillmair}}{2011}]{grillmair11}
{Grillmair} C.~J.,  2011, \mn@doi [\apj] {10.1088/0004-637X/738/1/98}, \href
  {http://adsabs.harvard.edu/abs/2011ApJ...738...98G} {738, 98}

\bibitem[\protect\citeauthoryear{{Hargis} et~al.,}{{Hargis}
  et~al.}{2016}]{hargis16}
{Hargis} J.~R.,  et~al., 2016, \mn@doi [\apj] {10.3847/0004-637X/818/1/39},
  \href {http://adsabs.harvard.edu/abs/2016ApJ...818...39H} {818, 39}

\bibitem[\protect\citeauthoryear{{Ibata}, {Irwin}, {Lewis}, {Ferguson}  \&
  {Tanvir}}{{Ibata} et~al.}{2003}]{ibata03}
{Ibata} R.~A.,  {Irwin} M.~J.,  {Lewis} G.~F.,  {Ferguson} A.~M.~N.,   {Tanvir}
  N.,  2003, \mn@doi [\mnras] {10.1046/j.1365-8711.2003.06545.x}, \href
  {http://adsabs.harvard.edu/abs/2003MNRAS.340L..21I} {340, L21}

\bibitem[\protect\citeauthoryear{{Ivezi{\'c}} et~al.,}{{Ivezi{\'c}}
  et~al.}{2008}]{ivezic08}
{Ivezi{\'c}} {\v Z}.,  et~al., 2008, \mn@doi [\apj] {10.1086/589678}, \href
  {http://adsabs.harvard.edu/abs/2008ApJ...684..287I} {684, 287}

\bibitem[\protect\citeauthoryear{{Lee} et~al.,}{{Lee} et~al.}{2008}]{lee08}
{Lee} Y.~S.,  et~al., 2008, \mn@doi [\aj] {10.1088/0004-6256/136/5/2022}, \href
  {http://adsabs.harvard.edu/abs/2008AJ....136.2022L} {136, 2022}

\bibitem[\protect\citeauthoryear{{Majewski}, {Ostheimer}, {Rocha-Pinto},
  {Patterson}, {Guhathakurta}  \& {Reitzel}}{{Majewski}
  et~al.}{2004}]{majewski04}
{Majewski} S.~R.,  {Ostheimer} J.~C.,  {Rocha-Pinto} H.~J.,  {Patterson} R.~J.,
   {Guhathakurta} P.,   {Reitzel} D.,  2004, \mn@doi [\apj] {10.1086/424586},
  \href {http://adsabs.harvard.edu/abs/2004ApJ...615..738M} {615, 738}

\bibitem[\protect\citeauthoryear{{Martin}, {Ibata}, {Bellazzini}, {Irwin},
  {Lewis}  \& {Dehnen}}{{Martin} et~al.}{2004}]{martin04}
{Martin} N.~F.,  {Ibata} R.~A.,  {Bellazzini} M.,  {Irwin} M.~J.,  {Lewis}
  G.~F.,   {Dehnen} W.,  2004, \mn@doi [\mnras]
  {10.1111/j.1365-2966.2004.07331.x}, \href
  {http://adsabs.harvard.edu/abs/2004MNRAS.348...12M} {348, 12}

\bibitem[\protect\citeauthoryear{{Meisner}, {Frebel}, {Juri{\'c}}  \&
  {Finkbeiner}}{{Meisner} et~al.}{2012}]{meisner12}
{Meisner} A.~M.,  {Frebel} A.,  {Juri{\'c}} M.,   {Finkbeiner} D.~P.,  2012,
  \mn@doi [\apj] {10.1088/0004-637X/753/2/116}, \href
  {http://adsabs.harvard.edu/abs/2012ApJ...753..116M} {753, 116}

\bibitem[\protect\citeauthoryear{{Newberg} et~al.,}{{Newberg}
  et~al.}{2002}]{newberg02}
{Newberg} H.~J.,  et~al., 2002, \mn@doi [\apj] {10.1086/338983}, \href
  {http://adsabs.harvard.edu/abs/2002ApJ...569..245N} {569, 245}

\bibitem[\protect\citeauthoryear{{Purcell}, {Bullock}  \&
  {Kazantzidis}}{{Purcell} et~al.}{2010}]{purcell10}
{Purcell} C.~W.,  {Bullock} J.~S.,   {Kazantzidis} S.,  2010, \mn@doi [\mnras]
  {10.1111/j.1365-2966.2010.16429.x}, \href
  {http://adsabs.harvard.edu/abs/2010MNRAS.404.1711P} {404, 1711}

\bibitem[\protect\citeauthoryear{{Reid} \& {Brunthaler}}{{Reid} \&
  {Brunthaler}}{2004}]{reid04}
{Reid} M.~J.,  {Brunthaler} A.,  2004, \mn@doi [\apj] {10.1086/424960}, \href
  {http://adsabs.harvard.edu/abs/2004ApJ...616..872R} {616, 872}

\bibitem[\protect\citeauthoryear{{Rocha-Pinto}, {Majewski}, {Skrutskie},
  {Crane}  \& {Patterson}}{{Rocha-Pinto} et~al.}{2004}]{rochapinto04}
{Rocha-Pinto} H.~J.,  {Majewski} S.~R.,  {Skrutskie} M.~F.,  {Crane} J.~D.,
  {Patterson} R.~J.,  2004, \mn@doi [\apj] {10.1086/424585}, \href
  {http://adsabs.harvard.edu/abs/2004ApJ...615..732R} {615, 732}

\bibitem[\protect\citeauthoryear{{Sch{\"o}nrich}, {Binney}  \&
  {Dehnen}}{{Sch{\"o}nrich} et~al.}{2010}]{schonrich10}
{Sch{\"o}nrich} R.,  {Binney} J.,   {Dehnen} W.,  2010, \mn@doi [\mnras]
  {10.1111/j.1365-2966.2010.16253.x}, \href
  {http://adsabs.harvard.edu/abs/2010MNRAS.403.1829S} {403, 1829}

\bibitem[\protect\citeauthoryear{{Sharma}, {Bland-Hawthorn}, {Johnston}  \&
  {Binney}}{{Sharma} et~al.}{2011}]{sharma11}
{Sharma} S.,  {Bland-Hawthorn} J.,  {Johnston} K.~V.,   {Binney} J.,  2011,
  \mn@doi [\apj] {10.1088/0004-637X/730/1/3}, \href
  {http://adsabs.harvard.edu/abs/2011ApJ...730....3S} {730, 3}

\bibitem[\protect\citeauthoryear{{Slater} et~al.,}{{Slater}
  et~al.}{2014}]{slater14}
{Slater} C.~T.,  et~al., 2014, \mn@doi [\apj] {10.1088/0004-637X/791/1/9},
  \href {http://adsabs.harvard.edu/abs/2014ApJ...791....9S} {791, 9}

\bibitem[\protect\citeauthoryear{{Velazquez} \& {White}}{{Velazquez} \&
  {White}}{1999}]{velazquez99}
{Velazquez} H.,  {White} S.~D.~M.,  1999, \mn@doi [\mnras]
  {10.1046/j.1365-8711.1999.02354.x}, \href
  {http://adsabs.harvard.edu/abs/1999MNRAS.304..254V} {304, 254}

\bibitem[\protect\citeauthoryear{{Xu}, {Newberg}, {Carlin}, {Liu}, {Deng},
  {Li}, {Sch{\"o}nrich}  \& {Yanny}}{{Xu} et~al.}{2015}]{xu15}
{Xu} Y.,  {Newberg} H.~J.,  {Carlin} J.~L.,  {Liu} C.,  {Deng} L.,  {Li} J.,
  {Sch{\"o}nrich} R.,   {Yanny} B.,  2015, \mn@doi [\apj]
  {10.1088/0004-637X/801/2/105}, \href
  {http://adsabs.harvard.edu/abs/2015ApJ...801..105X} {801, 105}

\bibitem[\protect\citeauthoryear{{Yanny} et~al.,}{{Yanny}
  et~al.}{2009}]{yanny09}
{Yanny} B.,  et~al., 2009, \mn@doi [\aj] {10.1088/0004-6256/137/5/4377}, \href
  {http://adsabs.harvard.edu/abs/2009AJ....137.4377Y} {137, 4377}

\bibitem[\protect\citeauthoryear{{York} et~al.,}{{York} et~al.}{2000}]{york00}
{York} D.~G.,  et~al., 2000, \mn@doi [\aj] {10.1086/301513}, \href
  {http://adsabs.harvard.edu/abs/2000AJ....120.1579Y} {120, 1579}

\makeatother
\end{thebibliography}

\bsp
\end{document}